# Full Poincaré polarimetry enabled through physical inference

CHAO HE[1,9], JIANYU LIN[2,3], JINTAO CHANG[4], JACOPO ANTONELLO[1], BEN DAI[5], JINGYU WANG[1], JIAHE CUI[1], JI QI[2,3], MIN WU[6], DANIEL S. ELSON[2,3], PENG XI[7], ANDREW FORBES[8] AND MARTIN J. BOOTH[1,10]

[1]*Department of Engineering Science, University of Oxford, Parks Road, Oxford, OX1 3PJ, UK*
[2]*Hamlyn Centre for Robotic Surgery, Imperial College London, London SW7 2AZ, UK*
[3]*Department of Surgery and Cancer, Imperial College London, London SW7 2AZ, UK*
[4]*Department of Physics, Tsinghua University, Beijing 100084, China*
[5]*Department of Statistics, The Chinese University of Hong Kong, Shatin, HK SAR, China*
[6]*Department of Computer Science, University of Oxford, Parks Road, Oxford, OX1 3QD, UK*
[7]*Department of Biomedical Engineering, College of Engineering, Peking University, Beijing 100871, China*
[8]*School of Physics, University of the Witwatersrand, Private Bag 3, Johannesburg 2050, South Africa*
[9]*chao.he@eng.ox.ac.uk*
[10]*martin.booth@eng.ox.ac.uk*

**Abstract:** *While polarisation sensing is vital in many areas of research, with applications spanning from microscopy to aerospace, traditional approaches are limited by method-related error amplification, accumulation and pre-processing steps, constraining the performance of single-shot polarimetry. Here, we put forward a new measurement paradigm to circumvent these limitations, based on the use of a universal full Poincaré generator to map all polarisation analyser states into a single vectorially structured light field. All vector components are analysed in a single-shot, extracting the vectorial state through inference from a physical model of the resulting image, providing a single-step sensing procedure. To demonstrate the feasibility of our approach, we use a common GRIN optic as our mapping device and show mean errors of <1% for each vector component. Our work paves the way for next-generation polarimetry, impacting a wide variety of applications that rely on vector measurement.*

## 1. Introduction

Polarisation sensing methods have wide applications which range from quantum physics to clinical applications [1–19]. They can be divided into two categories: time-resolved, where measurements are taken using a sequence of analysers in a time multiplexed manner, or single-shot, where different analysers are spatially multiplexed. Time resolved measurements can be easier to implement, but single-shot methods are crucial for applications with rapidly changing inputs or where high throughput is required. Single-shot polarimetry based on passive devices features advantages such as high stability compared with active counterparts. The standard measurement approaches in both methods are directly or indirectly related to a core measurement equation [20-24]: $S = inv(A) \cdot I$, where $S$ is the Stokes vector to be measured, and $I$ is the vector of intensities recorded at the detector. Matrix $A$ is known as the instrument matrix, which is determined by the system configuration. In order to enhance precision and accuracy, numerous attempts have been made to push $A$ towards an optimal matrix as it determines the properties of the error propagation and hence affects the performance [20-24]. An evaluation of systematic error amplification level can be performed using the condition number (CN) of $A$, with the theoretical minimum value [20-24] $\text{CN} = \sqrt{3}$. In practice, traditional approaches require various processes to be implemented before the matrix calculation, which includes denoising, optimisation, and calibration [20-33]. Although the detailed operating procedures are different among various polarimetry techniques, the complex error transfer and accumulation processes are fairly consistent [20-33] as all they essentially measure the vector property of the light beam in cumbersome and indirect ways. These place fundamental limits on precision and accuracy in single-shot polarimetry.

Here, we put forward a new paradigm that allows a single-shot polarimetry circumventing the error amplification factor and the necessary pre-processing steps. At the heart of our approach is the notion of Full Poincaré Beam (FPB) mapping, producing a particular form of vectorially structured light. The concept of a 'Full Poincaré beam' has been of great interest in optics for many years [34–39]. This interest stems from the uniqueness of the beam property, namely that the state of polarisation (SOP) of the beam's transverse cross section can cover the full Poincaré sphere, hence the name – FPB. Such beams are important for various applications utilising structured light, such as singularity analysis and beam shaping [34–39], and they have been harnessed in a wide range of polarimetric measurement techniques [42,40–48]. However, in existing FPB polarimetry, the error amplification and accumulation processes are still present, as these methods are based on traditional matrix based Stokes vector retrieval [31]. We harness a special class of the devices that can generate a FPB to enable our new paradigm, which we term a universal full Poincaré generator (UFPG). The essential feature here is the ability to generate different FPBs given different pure uniform incident SOPs (not all devices that can generate a FPB are also UFPGs; see Supplementary Note 1). The FPB generated via the UFPG provides all the information required to deduce the initial unknown SOP when mapped onto an intensity image through a polarisation analyser. We highlight here that the spatial variation in intensity provides a simple intuitive route to polarisation measurement, as there is a unique link between the intensity pattern and the input SOP. This provides new prospects for direct vector sensing through use of image processing to retrieve the incident SOP, rather than using an indirect matrix based method.

Figure 1 illustrates conceptually how the UFPG-based paradigm works, using a simple GRIN lens system as an example UFPG (other implementation options are explained in the discussion section). In this initial work, we focus on point measurements rather than imaging. A pinhole is used in our system in order to make sensing equivalent to a point measurement (see Supplementary Notes 1 and 2). We use an illumination beam with an arbitrarily chosen linear or elliptical SOP for demonstration purposes (in reality, this represents the unknown input state). A FPB is generated (Fig. 1b) after this passes through the GRIN lens [24, 35]. The output vector field is then filtered via a polarisation filter (PF) assembly (the right-hand circular (RC) SOP is selected as the eigenstate here) leading to a non-uniform intensity distribution (Fig. 1b) that can be recorded at the detector. Intuitively, one can understand that the brightest points within the intensity distribution must correspond to the eigenstate of the PF. The positions of these points depend upon the input state. Hence, the input SOP could – in principle – be read off directly from the positions of the brightest points, as long as the mapping of states to image position is known (see Fig. 1c). As the GRIN lens is a UFPG of order two, there are two points of maximum intensity (see Supplementary Notes 1 and 2). The nature of this UFPG sensing paradigm means that we have access, in principle, to a complete set of all possible analyser states in a single mapping. This goes beyond any other non-UFPG systems, as they cannot sense the 'complete' set of states in the same way and hence cannot enable the same paradigm; and also beyond other matrix calculation based UFPG polarimetry, as our paradigm retrieves the Stokes vector through physical inference from image analysis (see Supplementary Note 1). Our approach circumvents the above-mentioned mathematical limit, as here is no minimum error propagation amplification posed by the matrix process (see Supplementary Note 3). Instead, in theory its performance can be continuously boosted, even in a single-shot, by judicious choice of detector pixel size and imaging optics, and in practice further enhanced by the use of intelligence in the imaging, such as through machine learning.

Figure. 1d provides a visualized example of the link between the incident SOPs (shown on Poincaré sphere; Fig. 1d (i)), the SOPs related locations mapped on the intensity image (Fig. 1d (ii)), and a sketch of the pixel size on the detector (Fig. 1d (iii)). In order to determine the brightest point locations properly, the essential point of the new paradigm, we adopt a pure image-based sensing process (details in next section, and Supplementary Note 4). Instead of dividing the pipeline into denoising, optimisation, and intensity and polarisation calibration, we have managed to combine such processes in one step by remodelling the process into a combined fitting and estimation task based upon image processing and machine learning-enabled estimation of the brightest point. It is an optimal, system-based, integrated approach that enables high performance in precision, accuracy and stability.

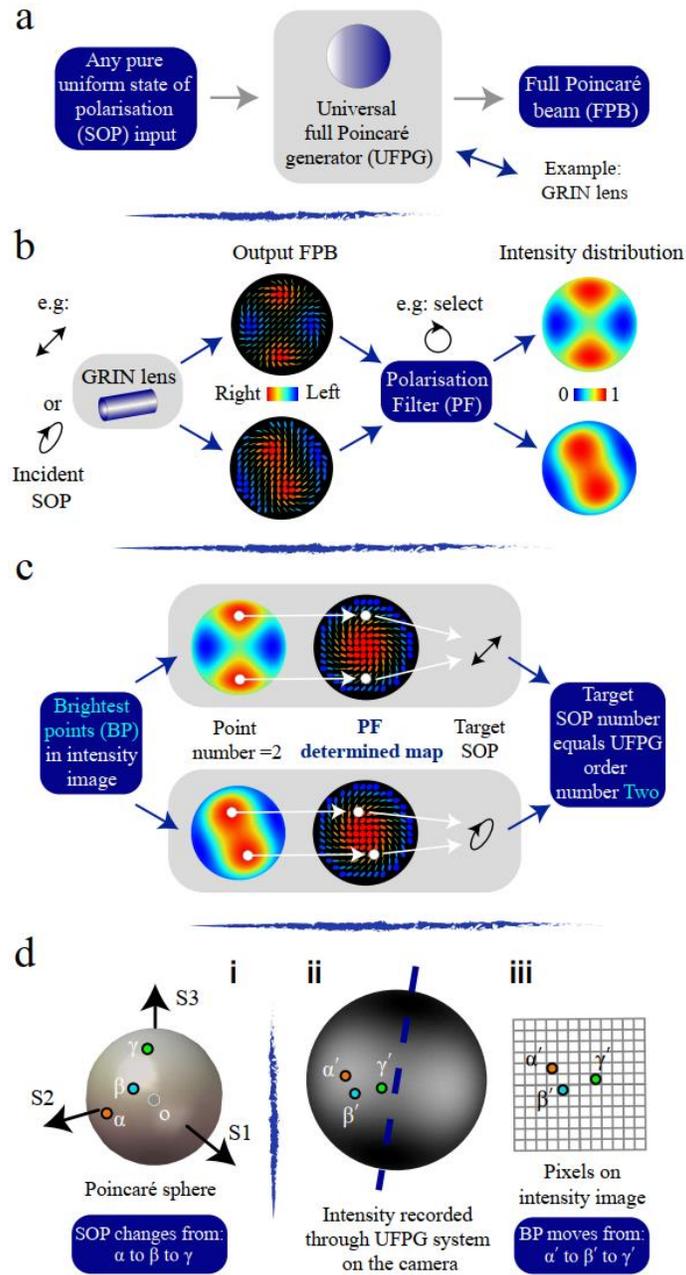

Fig. 1. Concepts and mechanisms of the UFPG paradigm. (a) Definition of a UFPG. (b) Intensity distribution generation process – including sending different incident SOPs through the GRIN lens system, obtaining corresponding output FPBs, and using a polarisation filter (PF) to create an intensity image. (c) Intuitive explanation of the measurement process, illustrated through finding brightest points (BP) on the intensity distribution and reading the input SOP from the locations corresponding to these points on the PF determined FPB map. In practice, all points in the image feed into the estimation process. (d) Mapping of SOPs between domains. (i) Example SOPs (points α, β, γ) on the Poincaré sphere. (ii) The corresponding points α′, β′, γ′, mapped from points α, β, γ on the acquired intensity image. (iii) A sketch of the points as pixel positions recorded on the detector. The pixel size acts as a theoretical limit on the sensitivity of the new paradigm (see Supplementary Note 4).

## 2. Results

The feasibly of such a UFPG polarimeter was tested experimentally, with the results shown in Figures 2 and 3. Effective realisation of this method consists of two steps. First, fitting of a model to the intensity data is required in order to estimate the position of the brightest points. This needs to be robust against temporal or spatial noise [20–30]. Secondly, the mapping between the detected positions and the related SOP should be properly identified. To implement these image processing steps, we drew upon machine-learning (ML) techniques [49–52], particularly through inspiration from the recent success of convolutional neural networks (CNN) in computer vision [49–52]. Importantly, rather than treating the complex polarisation measurement processes as separate steps as traditional methods have done [20–30], we implemented a single image-based 'end-to-end' solution [51,52]. This way we avoid the need for separate procedures for denoising, optimisation and calibration (which can each introduce separate errors).

An overview of the sensing process is shown in Fig.2a. It consists of three steps: 1) build a system as a polarisation state analyser using UFPG (see Supplementary Note 2 for details of the set-up); 2) train a neural network based on a predetermined UFPG system with different intensity images (including different brightest point locations) featuring known SOPs (see Supplementary Note 4 for details of the 'end-to-end' training); 3) record the experimental intensity images of unknown SOPs, load them into the image processing framework to determine the original SOPs, and output the prediction (see details in Supplementary Note 4). The core of the image processing framework – the neural network – is illustrated in Fig. 2b. An example of input experimental image, the constitution of the CNN, the output heatmap are also shown in the flow chart. Note here we use a widely-used/validated non-polarisation specific network [53] to solve polarisation measurement problem, based upon physical information. The rest of the framework (Fig. 2c) is an image post processing procedure, which links the brightest point locations and the SOPs via a look up table (LUT). For details refer to Supplementary Note 4. After such process is established, the online sensing process requires only step 3 to be repeated.

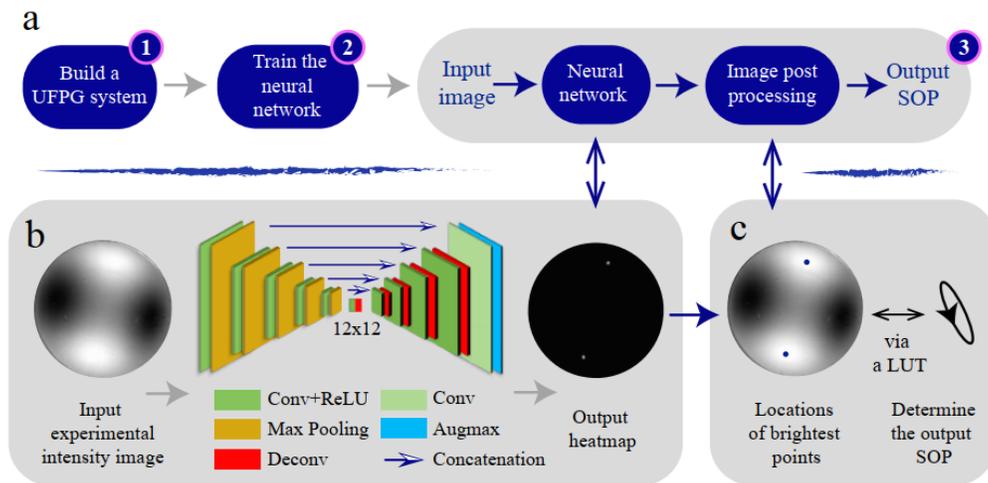

Fig. 2. An overview of the sensing process. (a) A flow chart of the ML-assisted UFPG based paradigm that includes three steps. (b) and (c) show one practical exemplar of using the CNN based image processing approach to retrieve the unknown incident SOP via an experimental intensity image. The neural network converts the input image to a mask featuring inferred positions of the brightest points (heatmap); while the image post processing locates the brightest

point locations and links to the output SOP via a LUP. Details are elaborated in Supplementary Note 4.

Different SOPs along three selected curves (a total of 900 points) on the Poincaré sphere were tested to validate the feasibility and for error analysis. Both theoretical and experimental data are demonstrated in Fig. 3a. Then 200 points were randomly sampled from the obtained data and the mean errors corresponding to elements of the Stokes vector: $S_1$, $S_2$, $S_3$, Euclidean distance, and degrees of polarisation (DOP) were calculated (Fig. 3b and 3c). We achieve exceptional measurement precision across the Poincaré sphere with stable performance and is robust compared to existing techniques – Fig. 3d, Supplementary Tables also give a quantitative comparison with several established methods to demonstrate the outstanding precision and accuracy of the new paradigm. Our results also confirm the ability to differentiate small scale polarisation changes in a single-shot with a sensitivity ~0.01 (this is defined mathematically in relation to the UFPG system property and camera pixel number; see details in Supplementary Note 5), around three times better than the well-established traditional single-shot point Stokes polarimeter, and more than six times better than other modern approaches based on light-matter interactions, such as plasmonics, spin-orbit effects and metasurfaces. In this work we mainly focus on fully polarized light detection and analysis. Detailed SOP retrieval under other conditions or with extra parameters such like DOP will form part of our future work.

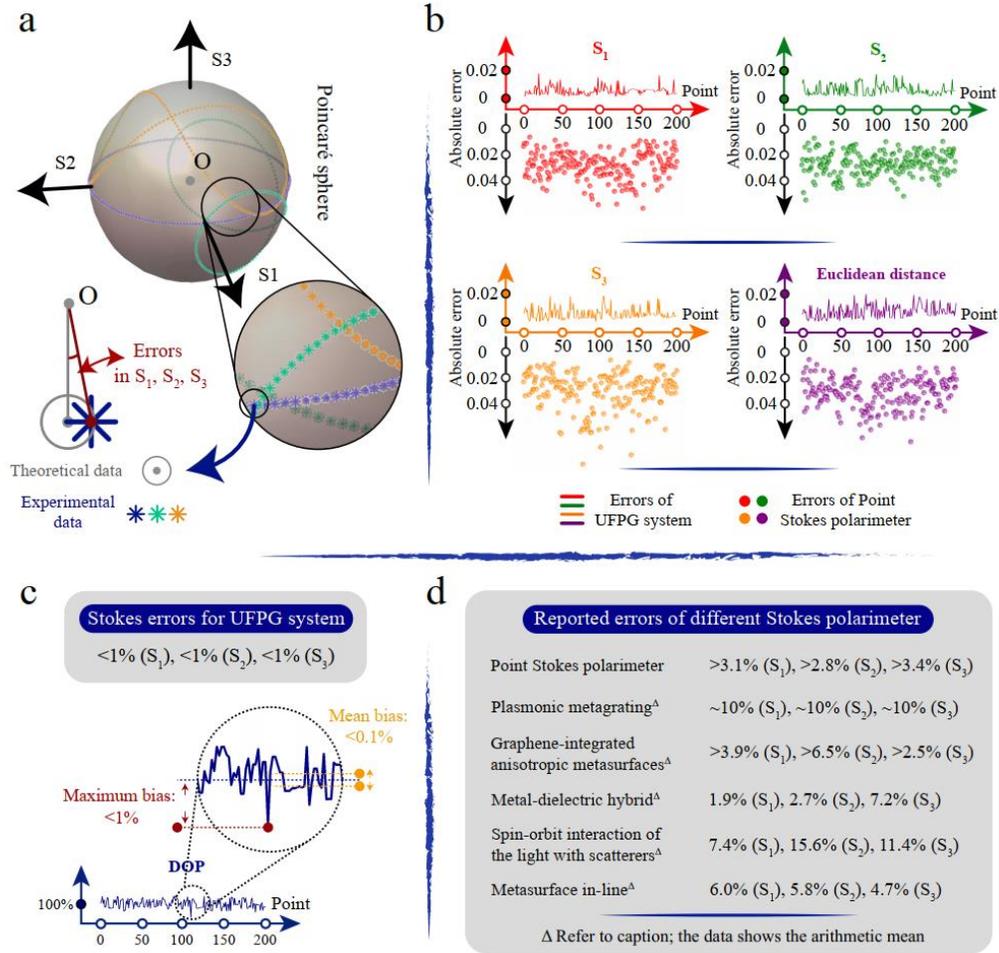

Fig. 3. Measurement performance verification of the UFPG paradigm, and comparison with recent reported techniques. (a) Three example curves (curves generated using an arbitrary state generator followed by a rotating quarter wave plate; For details of the generator see Supplementary Note 2) on the Poincaré sphere as a feasibility test for our new approach. Both theoretical and experimental data are presented on the sphere. Theoretical data are shown in grey circles; experimental counterparts are indicated by star symbols with different colours for different curves. The experimental set-up is described in Supplementary Note 2. (b) Precision evaluation process. The error diagrams for $S_1$, $S_2$, $S_3$, and *Euclidean distance* at 200 random points are acquired from the sampled curves on the Poincaré sphere. The same measurements were conducted by a traditional point Stokes polarimeter (i.e., a point measurement system; see Supplementary Note 5), whose measurement errors are plotted in coloured ball symbols for an intuitive comparison. (c) The errors of Stokes vector components and parameter *DOP* of the UFPG system are presented with calculated magnitudes. (d) The reported error performance of several modern polarimeters [54]. Compared to those existing counterparts [54–58], our new approach demonstrates the exceptional measurement precision and accuracy (see more data in Supplementary Tables 1 and 2).

In conclusion, we proposed a new paradigm for polarisation sensing using a novel mapping of any input SOP to a structured light field with a physical optic, which in subsequently processed by a machine learning algorithm. It featured exceptional measurement precision and accuracy in single-shot operation with robust performance. In this letter, we used a GRIN lens as the UFPG because it is non-pixelated, produces high beam quality, is low cost, and easily available [24]; however, the UFPG can also take the form of a stressed optical element [33,47,48], DMD, SLM, metasurface or an alternative device, as long as it is suitably configured. Further feasibility demonstrations are elaborated in Supplementary Note 5 and 6, including for low intensity measurement, cases of various DOP, as well as an example of weak polarisation error sensing. These validations show the solid performance of our new approach.

Future directions are available for further development of this concept. The sensitivity can be enhanced further by device substitution, as the fundamental minimum scale depends upon the number of pixels in the image (see Supplementary Note 4). The use of UFPGs with higher order numbers could possibly lead to greater robustness and precision, linked to the enhancement of the pattern complexity [51,52]. The depolarisation property can be readily extracted from the contrast of the intensity distribution image to reveal more polarimetric information (see Supplementary Note 6). Furthermore, the UFPG paradigm can be adapted to single-shot multi-point Stokes sensing (see Supplementary Note 7). There are obviously still challenges to overcome for this technique to be applied in certain future applications (such as 3D sensing), but our work opens a unique avenue towards precise sensing of weak vectorial information, which will provide useful capability in any application that relies upon polarisation sensing.

**Funding**

This project has received funding from the European Research Council (ERC) under the European Union's Horizon 2020 research and innovation programme (grant agreement n° 695140). C.H. would like to thank the support of the Junior Research Fellowship from St John's College, University of Oxford, and thank Mr. Binguo Chen from Tsinghua University for the useful discussions.

**Disclosures**

The authors declare no competing interests.

**Data availability statement**

Data underlying the results presented in this paper are not publicly available at this time but may be obtained from the authors upon reasonable request.

See Supplement information for supporting content.